\documentclass[onecolumn,fleqn,usenatbib]{article} 
\usepackage{hyperref}
\usepackage{grffile}
\usepackage{rotating}

\usepackage{xcolor}
\usepackage{newtxtext,newtxmath}
\usepackage{booktabs}
\usepackage{graphicx}
\usepackage{amsmath}

\usepackage{rotating}
\usepackage{multicol}
\usepackage{chngpage}
\usepackage{longtable}

\usepackage{mathtools}

\title{Constraining deceleration, jerk and transition redshift using cosmic chronometers, Type Ia supernovae and ISW effect.}
\date{2021}
\author{Syed Faisal ur Rahman\thanks{E-mail: faisalrahman36@hotmail.com}
	\\
Institute of Space and Planetary Astrophysics $(ISPA)$, \\University of Karachi $(UoK)$, Karachi, Pakistan\\
Karachi Institute of Technology and Entrepreneurship $(KITE)$, Karachi, Pakistan\\
Institute of Business Management $(IoBM)$,Karachi, Pakistan	
}
\begin{document}
	\maketitle
	\abstract{
		In this study we present constraints on the deceleration (q) and jerk (j) parameters using the late time integrated Sachs-Wolfe effect, type Ia supernovae, and H(z) data . We first directly measure the deceleration and jerk parameters using the cosmic chronometers data with the Taylor series expression of H(z).However, due to the  unusual variations in the deceleration parameter with slight changes in other parameters like  snap (s) and lerk (l), we found that direct measurements using the series expansion of the H(z) is not a suitable method for non-Lambda-CDM models and so we will need to derive the deceleration parameter after constraining density parameters and dark energy equation of state parameters. Then we present derived values of the deceleration parameter from Lambda CDM, WCDM and CPL models. We also discuss the transition redshift (zt) in relation with the deceleration parameter. Our best fit values for the deceleration parameter, after combining results from H(z), Union 2.1 and NVSS-ISW are obtained as -0.5808±0.025  for Lambda-CDM, -0.61±0.15 for both WCDM and CPL model. Our best fit for the combined jerk parameter for Lambda-CDM model is 1±3.971e-07, for WCDM model is 1.054±0.141 and for CPL model is 1.0654±0.1345. Also,the combined transition redshift is obtained as 0.724±0.047 for Lambda-CDM model.

	} 

	\section{Introduction}\label{introduction}
	
	Accelerated expansion of the universe and related concept of dark energy are among the most important concepts of modern cosmology. Cosmological studies before the discovery of the accelerated expansion of our universe were mainly focused at constraining the expansion rate of our universe via Hubble Constant and the deceleration parameter (q) \cite{Weinberg 2008} as then the matter dominated universe was assumed to be expanding with a decelerating rate. In this study we are constraining the deceleration parameter to see how this evolution evolves over the expansion history of universe. In an expanding universe, the deceleration parameter gives negative values which is observed by various observational results. While going back in time we can observe the transition of our universe from current accelerating phase to the deceleration phase in past at some redshift point known as transition redshift. We will also measure transition redshift by deriving it using matter and dark energy density parameters in $\Lambda$-CDM model and then compare it with transition redshift observed via the deceleration parameter under $\Lambda$-CDM, WCDM and CPL model assumptions.
	
	In an isotropic and homogeneous universe, we can start by defining the
	distance element as \cite{Weinberg 2008} \cite{Liddle 2003} \cite{Norton 2016}:
	
	\begin{equation}
	$\(ds^{2} = - dt^{2} + a(t)^{2}(\frac{dr^{2}}{1 - kr^{2}} + r^{2}d\Omega^{2})\)$
	\end{equation}

	Here a(t) is the scaling factor and k represent the spatial curvature.
	The ratio of the rate of change of the scale factor to the current value of
	scale factor is represented by the Hubble parameter:
	
	\begin{equation}
	$\(H(t) = \frac{\text{da}}{\text{adt}}\)$
	\end{equation}

	The deceleration parameter in relation with the Hubble parameter can be
	written as \cite{Rapetti et al. 2007} \cite{Weinberg 2008} \cite{Capozziello et al. 2019}:
	
	\begin{equation}
	$\(q(t) = - 1 - \frac{H'}{H^{2}}\)$
	\end{equation}
	
	Here H' is the time derivative of the Hubble parameter. We can write the
	deceleration parameter as a function of redshift in a flat universe as:
	
	\begin{equation}
	\label{qz}
	$\(q(z) = \Omega_{r}(z) + \frac{\Omega_{m}(z)}{2} + \Omega_{\Lambda}(z)\frac{1 + 3w(z)}{2}\)$
	\end{equation}
	
	Here \(\Omega_{r}\) is radiation density parameter, \(\Omega_{m}\) is the
	matter density parameter,\(\Omega_{\Lambda}\) is the dark energy density
	parameter and w(z) is the dark energy equation of state parameter which in
	\(\Lambda\)-CDM model is taken as a constant value of w=-1.
	
	\section{Measuring the deceleration parameter using H(z), ISW effect and type Ia supernovae data}\label{Deriving the deceleration parameter using H(z), ISW effect and type Ia supernovae data}
	
	\subsection{Cosmic Chronometers}
	
	In order to derive the results using equation \ref{qz}, we will first fit energy densities and dark energy equation of state (EoS) parameters using some available datasets. For this purpose we use direct H(z) measurement data which uses the cosmic chronometers approach  to study the expansion rate of the Universe as a function of redshift. As discussed in Moresco et al. 2018 \cite{Moresco et al. 2018} \cite{Moresco2016}, the expansion history of the Universe via the Hubble parameter as a function of redshift (z) can be constrained by measuring the differential age evolution of cosmic chronometers without assuming any particular cosmology. This cosmology independent way of measurement of H(z) evolution itself provides an interesting opportunity to test various models of cosmology quickly and more directly than some other methods.  
	Theoretically, for flat \(\Lambda\)-CDM model and extensions H(z) can be written as :
	
	\begin{equation}
	$\(H(z) = H_{0}\sqrt{\Omega_{\Lambda}I(z) + \Omega_{r}(1 + z)^{4} + \Omega_{m}(1 + z)^{3}}\)$
	\end{equation}

	Where, I(z) is defined as :
	\begin{equation}
	$\[I(z) = exp(3 \int_{0}^{z}{\left( \frac{1 + w_{\text{de}}\left( z^{'} \right)}{1 + z^{'}} \right)dz^{'}}\]$
	\end{equation}
	
	I(z) depends on the parametrization of the dark energy equation of state
	(EoS) and for standard \(\Lambda\)-CDM model with EoS as
	\(w_{\text{de}}\)(z)=-1 (constant), the multiplier I(z) becomes \(‘1’\)
	.
	
	Using the cosmic chronometer data, we can also measure the current
	deceleration parameter value at z=0 (q\textsubscript{0}) by using the
	Taylor expansion for H(y)=H\textsubscript{0}E(y).
	
	For this we can define a value y as \cite{Capozziello 2011} \cite{Rezaei 2020}:
	
	\[y = \frac{z}{z + 1}\]
	
	Now, we can define E(y) as:
	\begin{equation}
	E(y) =1 + k\textsubscript{1}y +
	(k\textsubscript{2}(y\textsuperscript{2})/2) +
	(k\textsubscript{3}(y\textsuperscript{3})/6) +
	(k\textsubscript{4}(y\textsuperscript{4})/24) 
	\end{equation}

	With,
	
	k\textsubscript{1} =1 + q\textsubscript{0}
	
	k\textsubscript{2} = 2 - (q\textsubscript{0}\textsuperscript{2}) +
	2q\textsubscript{0} + j\textsubscript{0}
	
	k\textsubscript{3} = 6 + 3(q\textsubscript{0}\textsuperscript{3}) -
	3(q\textsubscript{0}\textsuperscript{2}) + 6q\textsubscript{0} -
	4q\textsubscript{0}j\textsubscript{0} + 3j\textsubscript{0} -
	s\textsubscript{0}
	
	k\textsubscript{4} = -15(q\textsubscript{0}\textsuperscript{4}) +
	12(q\textsubscript{0}\textsuperscript{3}) +
	25(q\textsubscript{0}\textsuperscript{2})j\textsubscript{0} +
	7q\textsubscript{0}s\textsubscript{0} -
	4(j\textsubscript{0}\textsuperscript{2})
	-16q\textsubscript{0}j\textsubscript{0} -
	12(q\textsubscript{0}\textsuperscript{2}) + l\textsubscript{0} -
	4s\textsubscript{0} + 12j\textsubscript{0} + 24q\textsubscript{0} + 24
	
	\vspace{\baselineskip}
	For \(\Lambda\)-CDM case, we use
	j\textsubscript{0}=1,s\textsubscript{0}=0 and l\textsubscript{0}=0
	\cite{MuthukrishnaParkinson 2016}. For non-$\Lambda$ CDM cases, we use three configurations. In the first we one we fix s\textsubscript{0}=0 and l\textsubscript{0} as zero and keep H\textsubscript{0}, q\textsubscript{0} and j\textsubscript{0} free. In the second non-$\Lambda$ CDM case, we fix H\textsubscript{0} to $\Lambda$-CDM result, and s\textsubscript{0}=0 and l\textsubscript{0}=0. In the third case, we only fix H\textsubscript{0} and use other parameters as free. Table   \ref{tab:table_1} presents the results for different models. For our study we use boundary conditions\(\ 65 \leq H_{0} \leq 75\), \(\ -5 \leq q_{0} \leq 5\), \(\ -10 \leq j_{0} \leq 10\), \(\ -500 \leq s_{0} \leq 500\) and \(\ -2000 \leq l_{0} \leq 2000\).

	\begin{table}[htbp]
		\centering
		\caption{Measurement of q\textsubscript{0}, j\textsubscript{0}, l\textsubscript{0} and s\textsubscript{0} from Taylor series expression by using Moresco et al., 2016 data \cite{Moresco2016} with H(y=\(\frac{\mathbf{z}}{\mathbf{z + 1}}\) from  Capozziello et al., 2011 \cite{Capozziello 2011} and  Rezaei et al. 2020 \cite{Rezaei 2020}.}
		\begin{tabular}{lrlllll}
			\toprule
			\textbf{Model\textbackslash{}Parameters} &         & \textbf{H\textsubscript{0}} & \textbf{q\textsubscript{0}} & \textbf{j\textsubscript{0}} & \textbf{s\textsubscript{0}} & \textbf{l\textsubscript{0}} \\
			\midrule
			$\Lambda$-CDM &         & 68.4±3.1 & -0.34±0.24 & 1(fixed) & 0(fixed) & 0(fixed) \\
			Non-Lambda-CDM-I &         & 70.1±3.2 & -0.82±0.26 & 2.46±0.63 & 0(fixed) & 0(fixed) \\
			Non-Lambda-CDM-II &         & 68.4(fixed) & -0.72±0.19 & 2.32±0.64 & 0(fixed) & 0(fixed) \\
			Non-Lambda-CDM-III &         & 68.4(fixed) & -0.45±0.48 & -0.8±6.6 & -9±90   & 210±740 \\
		\end{tabular}%
		\label{tab:table_1}%
	\end{table}%
	
	
	We use H(z) data set provided by Moresco et al., 2016 \cite{Moresco2016} \cite{Brinckmann et al. 2019} for our cosmic chronometer
	related analysis. We an see in table \ref{tab:table_1} that Taylor series expansion results show extremely large uncertainties if we use snap and lerk as free parameters. Also, they don't give us much idea about the underlying cosmology. Therefore, we will use the derived deceleration and jerk parameters for our further analysis. Table \ref{tab:table_2} presents the results for the derived deceleration and jerk parameters using the cosmic chronometers or H(z) data.
	
	\subsection{Type Ia supernovae}
	
	Type Ia supernovae have been a useful tool in constraining cosmological
	parameters especially in context of the accelerated expansion of our
	universe. We can write the distance modulus for as a difference of
	apparent magnitude (m) and absolute magnitude (M) of the type Ia
	supernovae as \cite{Davis 2012} \cite{Suzuki et al. 2012} \cite{Amanullah et al. 2010}:
	
	\[\mu = m - M\]
	
	Also, luminosity distance and distance modulus are related as:
	
	\begin{equation}
	\label{eq_muz}
	$\(\mu(z) = 5log\lbrack DL(z)\rbrack + 25\)$
	\end{equation}

	We can get apparent magnitude in the form:
	\begin{equation}
	\label{eq_mz}
	$\(m(z) = 5log(DL(z)) + M + 25\)$
	\end{equation}
	
	With the inclusion of observational factors like color (k), shape(s) and
	the probability that the supernova belongs in the low-host-mass category
	(P), equation \ref{eq_mz} becomes :
	
	\begin{equation}
	\label{eq_mz_2}
	$$ m(z) = 5log(DL(z)) + M - \alpha s + \beta k - \delta P + 25 $$
	\end{equation}
	
	We use Union 2.1 distance modulus dataset which is publicly shared by
	Supernova Cosmology Project (SCP). The dataset is comprised of 580 type
	Ia supernovae which passed the usability cuts. The dataset is comprised
	of redshift range \(0.015 \leq z \leq 1.414\) with median redshift at
	\(z \approx 0.294\). Due to the degeneracy issue between Hubble Constant
	(H\textsubscript{0}) and the absolute magnitude of type Ia supernovae
	(M), we can separate contribution
	of H\textsubscript{0} and uncertainty in the absolute magnitude 'M' from equations \ref{eq_muz} and \ref{eq_mz}, as \cite{Davis 2012} :
	
	\begin{equation}
	$$ M' = 25 + 5log( \frac{c}{H_{0}}) +  \sigma_{M} $$
	\end{equation}
	
	We can then marginalize over this part to constrain densities and EoS
	parameters. We use Union 2.1 data set for our type Ia supernovae related analysis \cite{Suzuki et al. 2012}.
	
	\subsection{ISW Effect}
	The late time integrated Sachs-Wolfe effect (ISW) deals with the blue-shifting and red-shifting of the CMB photons due to the presence of large scale structures and super-voids respectively (please read: \cite{SachsWolfe 1967} \cite{Afshordi 2004} \cite{RahmanIqbal 2019} \cite{Hojjate et al. 2011} \cite{Laureijis et al. 2011} \cite{Loverde 2008} \cite{Nolta et al. 2004} \cite{Vagnozzi2020}). The cross-correlation angular power spectrum
	coefficient `Cl' can be calculated as:
	
	\begin{equation}
	$$Cl_{\text{gt}} = 4\pi\int\frac{\text{dk}}{k}\Delta^{2}(k)Wl_{g}(k)Wl_{t}(k)$$
	\end{equation}

	Here, \(\text{Wl}_{g}(k)\text{\ and\ W}l_{t}(k)\) are galaxy and
	temperature window functions respectively and $\Delta^{2}(k)$  is the
	logarithmic matter power spectrum. Galaxy window function depends on
	redshift distribution (dN/dz) and galaxy bias (b).We use redshift
	distribution as in dataset for NVSS-ISW provided by Stölzner et al. 2018 \cite{Stolzner et al. 2018}, Brinckmann \& Lesgourgues 2019 \cite{Brinckmann et al. 2019} and Audren et al. 2013  \cite{Audren et al. 2013} which provide CMB-galaxy cross-correlation angular power spectrum for Planck 2015 \cite{Planck 2016} \cite{Planck 2016b}  and NVSS \cite{Condon1998} \cite{Blake2004}. We also use redshift dependent galaxy bias for NVSS \cite{Planck 2016b} \cite{Planck 2016}.
	
	\begin{equation}
	$$b(z) = 0.90 [1 + 0.54(1 + z)^{2}]$$
	\end{equation}
	
	Another factor which affects the galaxy window function and so the ISW effect is the magnification bias due to gravitational lensing. Magnification bias is dependent on the slope ‘$\alpha$’ of the integral count, $N(>S)=CS^{-\alpha}$. However as discussed in \cite{Bianchini2015} \cite{Loverde 2007}, ‘$\alpha$’ only plays its part when it is greater or less than ‘1’. In NVSS integral count’s case, ‘$\alpha$’ is almost equal to ‘1’ and so did not significantly affect our theoretical calculations.
	
	For calculations related to the minimum \(\chi^{2}\) and mean likelihood, we use
	diagonal of the covariance matrix provided by Stölzner et al. 2018 \cite{Stolzner et al. 2018}, Brinckmann \& Lesgourgues 2019 \cite{Brinckmann et al. 2019} and Audren et al. 2013  \cite{Audren et al. 2013}. We use
	cross-correlation Cl values from multipole (l)=10 to 100 as for higher
	multipole ranges CMB lensing and Sunyaev--Zeldovich effect play a
	significant part in the overall CMB anisotropy power spectrum \cite{SunyaevZeldovich 1972} \cite{ShajibWright 2016}. Apart from cosmology parameters, we also fit a parameter AISW which quantifies ISW amplitude
	as in Stölzner et al. 2018 \cite{Stolzner et al. 2018} and also to deal with possible effect on
	'Cl' values due to error in modeling the galaxy bias, b(z). AISW is not equal to 1 can
	either mean disagreement with \(\Lambda\)-CDM or modeling issues with
	the galaxy bias. To calculate AISW, we use:
	
	\begin{equation}
	\chi^2=\sum((Cl_{th}-AISW.Cl_{obs})/(\sigma^2))
	\end{equation}
	
	\section{Likelihood}
	We calculated likelihood (L) as:
	\begin{equation}
	$$L = \text{exp}\frac{- \Delta\chi^{2}}{2}$$
	\end{equation}
	Here, \(\Delta\chi^{2}\) is the chi-square minus the minimum chi-square value for the parameter sets being tested by using the theoretical models and data \footnote{Sample code for likelihood calculations using H(z) data: \url{https://github.com/faisalrahman36/Workshop_labs/tree/master/Workshop_2_Cosmology_examples/Hz_Data}}. For parameter fitting, we use mean likelihood instead of maximum
	likelihood in order to minimize the effect of parameter boundary cuts \cite{Davis 2012}.
	
	For our study we use boundary conditions\(\ 65 \leq H_{0} \leq 75\),
	\(0.6 \leq \ \mathrm{\Omega}_{\Lambda}\  \leq 0.8,\  - 1.5 \leq w_{0} \leq - \ 1/3\ \)and\(\  - 2 \leq w_{a} \leq 2\).
	We are assuming flatness with all the models for our analysis.
	
	We use mean likelihood analysis to constrain parameters \cite{Davis 2012}
	with \(\Lambda\)-CDM, WCDM and CPL models \cite{ChevallierPolarski 2001} \cite{Linder 2003}. Figures (\ref{fig:qzhz} \ref{fig:qzunion2p1} \ref{fig:qziswnvss}) show deceleration
	parameter evolution over different redshifts from H(z),NVSS-ISW, and
	Union 2.1 for \(\Lambda\)-CDM, WCDM and CPL models and their best fit
	values can be seen in table \ref{tab:table_2}.
	
	The results are given in \ref{tab:table_2}. We can see that for\(\Lambda\)-CDM
	model, results for the deceleration parameter (q\textsubscript{0})
	derived from densities and EoS parameters using H(z), ISW and Union 2.1
	type Ia supernovae are not only consistent with each other within one
	standard deviation but are also in agreement with table \ref{tab:table_1}, within one
	standard deviation.
	
	However, results for WCDM and CPL parameters, derived from H(z), ISW and
	Union 2.1 type Ia supernovae, are quite different from non
	\(\Lambda\)-CDM q\textsubscript{0} results from table \ref{tab:table_1}. This is
	mainly because of the limitations with Taylor series luminosity distance
	form used for measurements in table \ref{tab:table_1}. It is also worth noting that
	all $w_{0}$ results from WCDM and CPL using H(z), ISW and Union 2.1 type Ia
	supernovae are not too different from our Lambda-CDM assumption of a
	constant equation of EoS parameter, w=-1. They all agree with w=-1,
	within one standard deviation despite being different in nature and
	redshift ranges. Even though we kept phantom energy region
	(w\textless-1) in our boundary conditions, but our results still agree
	with a likely constant w=-1. Phantom energy is theoretical concept
	associated with the acceleration of our universe where w\textless-1 can
	lead to the possibility of a "Big Rip" fate of our universe, according
	to which all matter from large scale structures to subatomic particles
	getting torn apart \cite{Farnes2018} \cite{Vikman2005}. Vagnozzi et al. 2018 \cite{Vagnozzi2018} suggested that quintessence dark energy models can be ruled out in coming years, independently of cosmological observations if long-baseline neutrino experiments measure the neutrino mass ordering to be inverted. However, this will require confirmation from cosmological studies using other signatures as well.
	
	Another interesting thing are the values of Hubble constant
	(H\textsubscript{0}),in H(z) and ISW from NVSS, are in agreement with
	each other. The mean values are closer to 70 as observed by the
	gravitational waves and H\textsubscript{0} based on a calibration of the
	Tip of the Red Giant Branch (TRGB) applied to type Ia supernovae (SNeIa)
	\cite{Freedman et al. 2019} \cite{LIGO 2017} \cite{Abbott et al. 2016} \cite{Abbott et al. 2017} and
	standard deviation margins bring them in the range of the extreme values
	observed by Planck's CMB measurements, Dark Energy Survey (DES),
	H0LiCOW's combined results \cite{Macaulay et al. 2019} \cite{Birrer et al. 2018} \cite{Riess et al. 2019}. There are still other challenges associated with the observational signatures from which the datasets are taken,
	apart from disagreements over the expansion rate of our universe. There
	are issues like the cosmic cold spot or the CMB cold spot \cite{Rudnick et al. 2007} \cite{Rahman 2020} which isn't agreeing with the ISW effect results based on current observations. This can be due to issues related
	to galaxy bias, magnification bias, galaxy classification, redshift
	distribution and other issues involved in understanding the galaxy
	window function and galaxy-CMB maps cross-correlations \cite{Nolta et al. 2004} \cite{Raccanelli2008} \cite{Raccanelli2012}. There are also challenges associated with the type Ia supernovae observations which is
	attracting some attention mainly due to the expansion rate debate.
	
	There are also possibilities of such issues arising due to our lack of
	understanding of things like curvature \cite{Valentino et al. 2020} \cite{Vagnozzi2021a} \cite{Vagnozzi2021b} \cite{Handley2021} or nature of inflation in different areas \cite{Wang et. al. 2016} or some other
	exotic phenomenon \cite{Rahman 2020}.
	
	\section{Transition redshift}\label{transition-redshift}
	
	A useful quantity to study the accelerated expansion of our universe is
	to measure the transition redshift (zt). The transition redshift can be
	defined as the redshift at which universe enters into the accelerating
	phase from an earlier decelerating phase. In flat Lambda-CDM models, we
	can write zt as \cite{Moresco2016}:
	
	\begin{equation}
	$$zt = \frac{2\Omega_{\Lambda}}{\Omega_{m}}^{1/3} - 1$$
	\end{equation}
	
\begin{figure}
		\centering
		\includegraphics[width=1\linewidth]{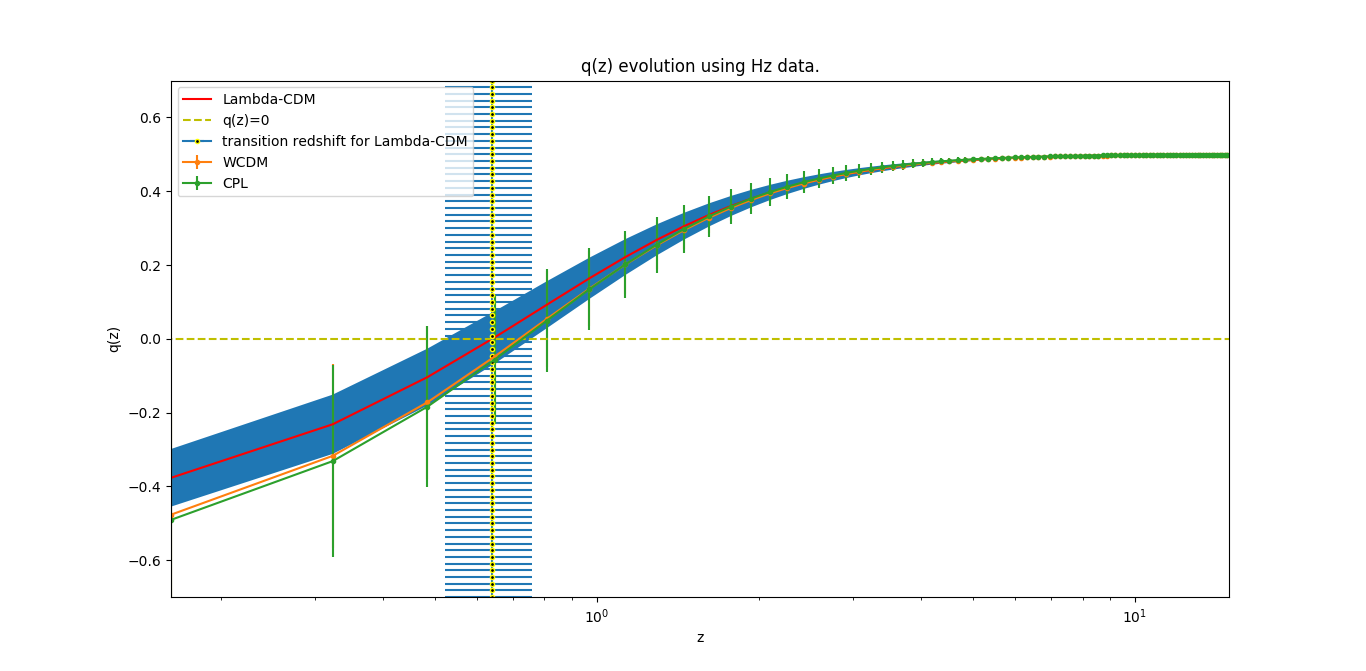}
		\caption[Deceleration parameter and transition redshift measurements from H(z) data.]{Deceleration parameter and transition redshift measurements from H(z) data.}
		\label{fig:qzhz}
	\end{figure}

\begin{figure}
		\centering
		\includegraphics[width=1\linewidth]{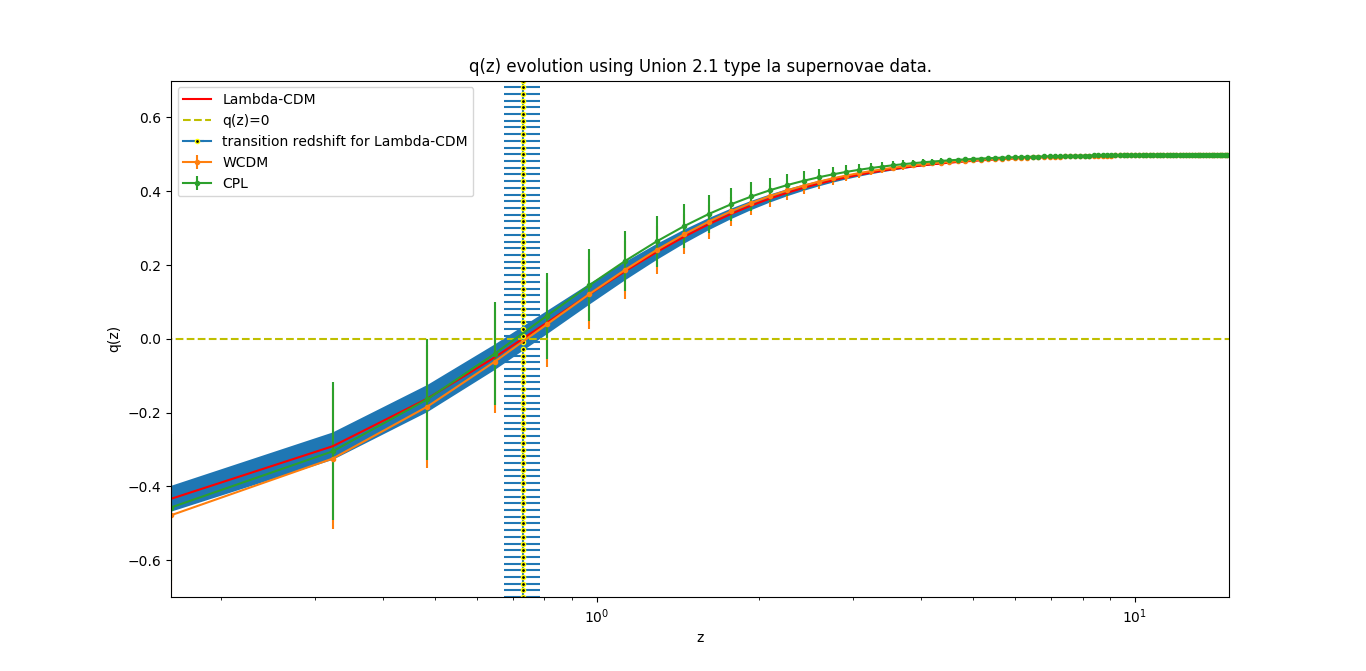}
		\caption[Deceleration parameter and transition redshift from Union 2.1 type Ia supernovae data.]{Deceleration parameter and transition redshift from Union 2.1 type Ia supernovae data.}
		\label{fig:qzunion2p1}
	\end{figure}

\begin{figure}
		\centering
		\includegraphics[width=1\linewidth]{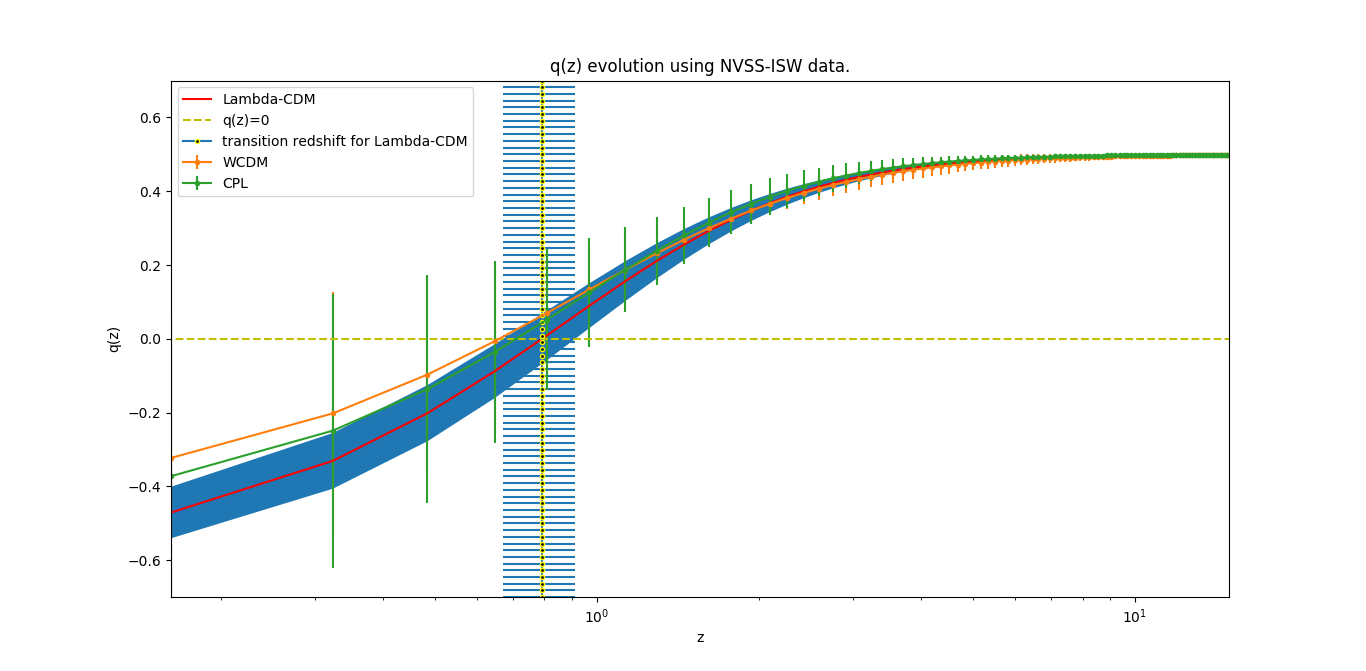}
		\caption[Deceleration parameter and transition redshift from NVSS-ISW data.]{Deceleration parameter and transition redshift from NVSS-ISW data.}
		\label{fig:qziswnvss}
	\end{figure}

	Table \ref{tab:table_3} gives the results we get from H(z),NVSS-ISW, and Union 2.1 for
	flat Lambda-CDM model. The results are consistent within one standard
	deviation, with the results obtained from previous studies using CMB
	anisotropies, BAO, type Ia supernovae and cosmic chronometers for
	\(\Lambda\)-CDM model \cite{Moresco2016} \cite{Busca et al. 2013} \cite{Lima et al. 2012} \cite{Hinshaw et al. 2013} \cite{Planck 2018}. However, our best fit transition redshift
	results are slightly larger than the measurements from cosmological
	model independent approach discussed in Moresco et al. 2016 \cite{Moresco2016}. Figures
	(\ref{fig:qzhz} \ref{fig:qzunion2p1} \ref{fig:qziswnvss}) show transition redshifts from H(z),NVSS-ISW, and
	Union 2.1 for flat \(\Lambda\)-CDM model and their best fit values can
	be seen in table \ref{tab:table_3}.

	%

	%
	%

	\section{Jerk parameter} 
	Another quantity which is helpful in understanding the acceleration of the universe especially its deviations from the $\Lambda$-CDM is the jerk parameter. Jerk parameter can be helpful in understanding the transitions between phases of different cosmic accelerations. 
	The jerk parameter, j \cite{Weinberg 2008} \cite{Blandford2004} \cite{MamonDas2017}, is a dimensionless quantity obtained by taking the third derivative of the scale factor ‘a(t)’ w.r.t cosmic time \cite{Blandford2004}.  We can calculate the jerk parameter as a function of redshift, using the deceleration parameter, q as \cite{MamonDas2017}:
	
	\begin{equation}
	$$j(z)=  [q(z)(2q(z)+ 1)+ \frac{dq}{dz}  (1 + z)  ]$$
	\end{equation}

	\begin{figure}
		\centering
		\includegraphics[width=1\linewidth]{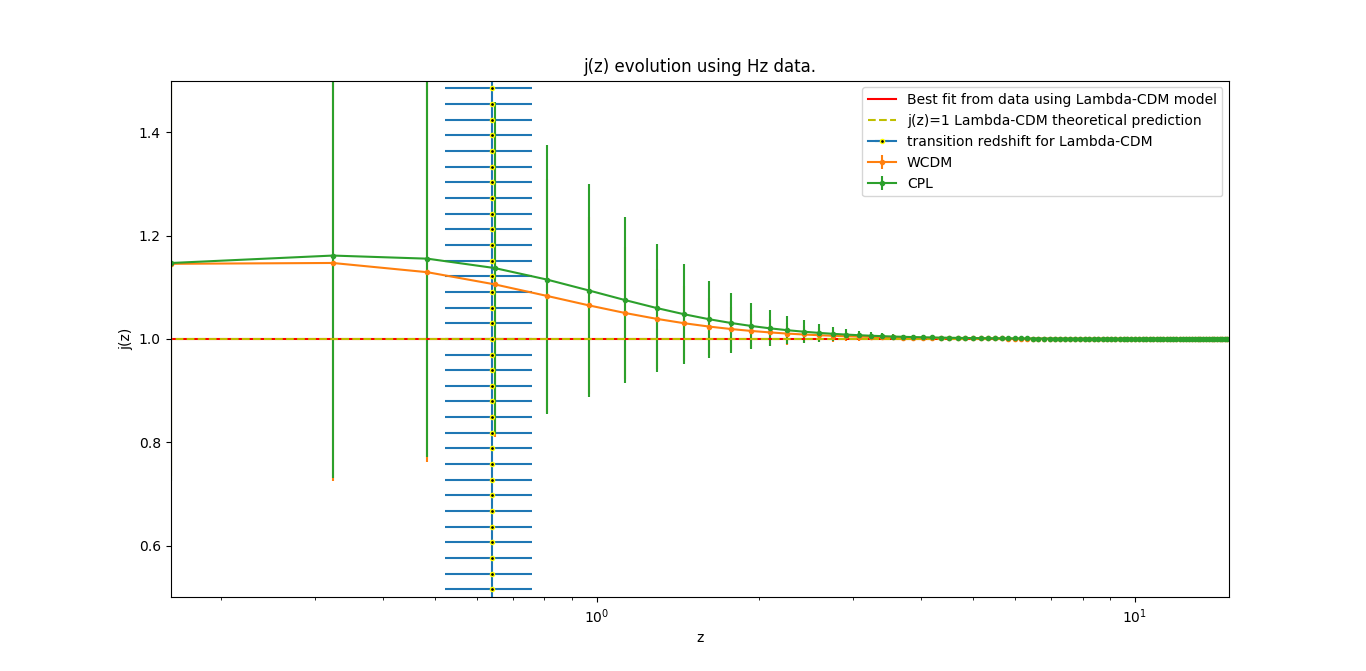}
		\caption{Jerk parameter and transition redshift measurements from H(z) data.}
		\label{fig:jzhz}
	\end{figure}

	\begin{figure}
		\centering
		\includegraphics[width=1\linewidth]{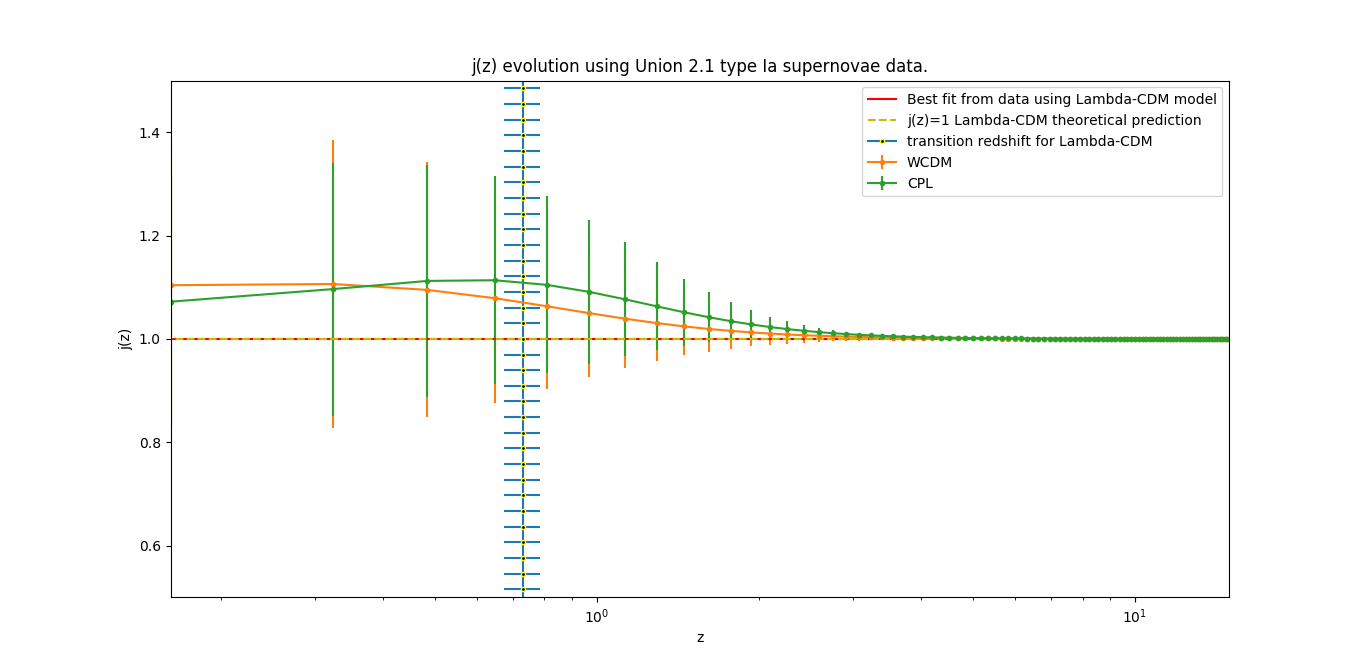}
		\caption{Jerk parameter and transition redshift from Union 2.1 type Ia supernovae data.}
		\label{fig:jzunion2p1}
	\end{figure}

	\begin{figure}
		\centering
		\includegraphics[width=1\linewidth]{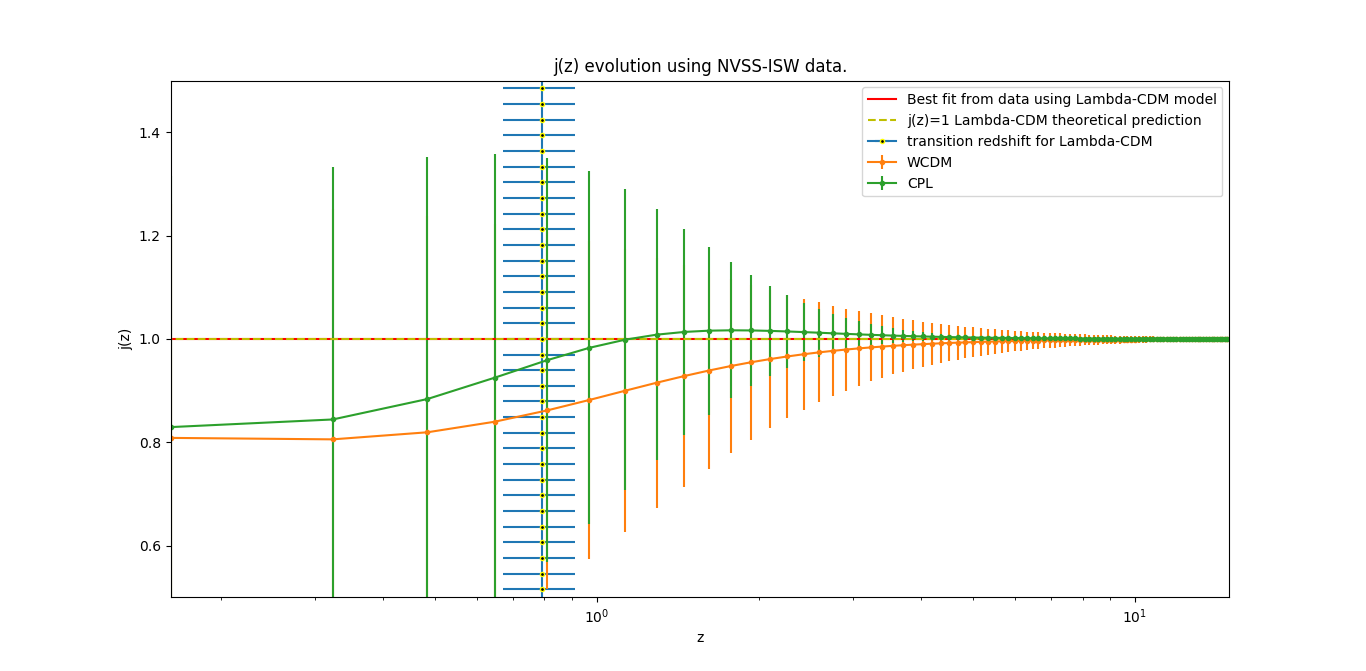}
		\caption{Jerk parameter and transition redshift from NVSS-ISW data.}
		\label{fig:jziswnvss}
	\end{figure}

	Figures (\ref{fig:jzhz} \ref{fig:jzunion2p1} \ref{fig:jziswnvss}) show the evolution of jerk parameter as a function of redshift. We can see that they agree with the $\Lambda$-CDM assumption of j\textsubscript{0}=1 for all models and data. In the figures (\ref{fig:jzhz} \ref{fig:jzunion2p1} \ref{fig:jziswnvss}), we plot both j(z)=1 assumption and j(z) evolution from the data for $\Lambda$-CDM along with WCDM and CPL models. Similar can be seen in tables \ref{tab:table_2} and \ref{tab:table_4} for the current j\textsubscript{0} values using  H(z), Union 2.1 and ISW-NVSS datasets with $\Lambda$-CDM, WCDM and CPL models.

	\section{Combined constraints}\label{combined-constraints}
	
	In order to obtain combined constraints for the deceleration parameter
	(q\textsubscript{0}), jerk parameter (j\textsubscript{0}) and transition redshift (zt), we use
	inverse-variance weighted average mean. We first combine lower redshift
	signatures from Union 2.1 and H(z), then we combine both higher redshift
	ISW signatures and then we combine results from all datasets in our study.
	We can see in table \ref{tab:table_4} that results agree with each other within one
	standard deviation. However in case of both (q\textsubscript{0}) and (j\textsubscript{0}),  deviations for WCDM and CPL are
	relatively higher especially for high redshift ISW studies. However,
	standard deviations improved when all datasets are combined together.
	For \(\Lambda\) -CDM case, results from tables \ref{tab:table_1} \ref{tab:table_2} and \ref{tab:table_4} agree with
	each other within one standard deviation but for non-\(\Lambda\) CDM cases, we can see some
	disagreement in the jerk parameter which is likely due to the greater uncertainties involved with the
	H(z)'s Taylor series expression approximations when we free jerk, snap and lerk parameters. 
	
	We can also see that results for transition redshift in table \ref{tab:table_5} giving a value $\approx$0.724, giving slightly lesser standard deviations than table \ref{tab:table_3}. The combined transition redshift is closed to previous \(\Lambda\) -CDM based estimates  \cite{Moresco2016} \cite{Busca et al. 2013} \cite{Lima et al. 2012} \cite{Hinshaw et al. 2013} \cite{Planck 2018} but higher than the cosmological model independent method results presented in Moresco et al. 2016 \cite{Moresco2016}.

	
	\begin{sidewaystable}
		
		\centering
		\caption{Cosmological parameters and derived deceleration parameters from H(z),NVSS-ISW and Union 2.1.}
		\begin{tabular}{llllllll}
			\toprule
			\multicolumn{1}{|l}{\textbf{Dataset/Model}} & \textbf{AISW} & \textbf{H0} & \textbf{Omega\_Lambda} & \textbf{w0} & \textbf{wa} & \textbf{q0 (derived)} & \textbf{j0 (derived)} \\
			\midrule
			\textbf{H(z)} &         &         &         & \textcolor[rgb]{ .133,  .133,  .133}{} & \textcolor[rgb]{ .133,  .133,  .133}{} &         &  \\
			Lambda-CDM & N/A     & 68.8±2.4 & 0.688±0.046 &         &         & -0.532±0.069 & 1± 3.4786e-06 \\
			WCDM    & N/A     & 69.5±3  & 0.688±0.048 & -1.1±0.26 &         & -0.6352±0.28 & 1.1±0.2684 \\
			CPL     & N/A     & 69.7±3  & 0.689±0.053 & -1.11±0.26 & -0.2±1.2 & -0.647±0.282 & 1.1137±0.269 \\
			&         &         &         &         &         &         &  \\
			\textbf{NVSS-ISW} &         &         &         &         &         &         &  \\
			Lambda-CDM & 2.92±0.79 & 70.2±3.4 & 0.742±0.039 &         &         & -0.613±0.0585 & 1±4.633e-06 \\
			WCDM    & 2.93±0.78 & 70.2 (fixed from LCDM) & 0.743±0.039 & -0.86±0.37 &         & -0.458±0.415 & 0.844±0.41245 \\
			CPL     & 2.93 (fixed from WCDM) & 70.2 (fixed from LCDM) & 0.746±0.036 & -0.9±0.39 & -0.4±1.1 & -0.507±0.439 & 0.888±0.4364 \\
			&         &         &         &         &         &         &  \\
			\textbf{Union 2.1} &         &         &         &         &         &         &  \\
			Lambda-CDM & N/A     & Marginalized & 0.721±0.02 &         &         & -0.5815±0.03 & 1±4.012e-07 \\
			WCDM    & N/A     & Marginalized & 0.705±0.055 & -1.07±0.17 &         & -0.6315±0.2 & 1.074±0.18 \\
			CPL     & N/A     & Marginalized & 0.692±0.059 & -1.07±0.16 & -0.4±1  & -0.61066±0.19118 & 1.07266±0.1662 \\
		\end{tabular}%
		\label{tab:table_2}%
	\end{sidewaystable}
	%
	%
	\begin{table}[htbp]
		\centering
		\caption{Transition redshift(zt) from H(z), NVSS-ISW and Union 2.1 for flat Lambda-CDM model.}
		\begin{tabular}{ll}
			\toprule
			\textbf{Dataset} & \textbf{Tranision Redshift (zt)} \\
			\midrule
			&  \\
			H(z)    & 0.639915± 0.11714 \\
			NVSS-ISW & 0.792 ± 0.122 \\
			Union 2.1 & 0.728968  ± 0.0573 \\
		\end{tabular}%
		\label{tab:table_3}%
	\end{table}%
	
	%
	\begin{table}[htbp]
		\centering
		\caption{Combined deceleration and jerk parameters from H(z),NVSS-ISW and Union 2.1}
		\begin{tabular}{rlll}
			\toprule
			\multicolumn{1}{l}{\textbf{Model}} & \textbf{Datasets} & \textbf{q0} & \textbf{j0} \\
			\midrule
			&         &         &  \\
			\multicolumn{1}{l}{Lambda-CDM} &         &         &  \\
			& H(z)+Union 2.1 & -0.574±0.0275 & 1±3.986e-07 \\
			& H(z)+Union 2.1+NVSS-ISW & -0.5808±0.025 & 1±3.971e-07 \\
			&         &         &  \\
			\multicolumn{1}{l}{WCDM} &         &         &  \\
			& H(z)+Union 2.1 & -0.633±0.163 & 1.082±0.1495 \\
			& H(z)+Union 2.1+NVSS-ISW & -0.61±0.15 & 1.054±0.141 \\
			&         &         &  \\
			\multicolumn{1}{l}{CPL} &         &         &  \\
			& H(z)+Union 2.1 & -0.622±0.158 & 1.084±0.1414 \\
			& H(z)+Union 2.1+NVSS-ISW & -0.61±0.15 & 1.0654±0.1345 \\
		\end{tabular}%
		\label{tab:table_4}%
	\end{table}%
	
	%
	%
	\begin{table}[htbp]
		\centering
		\caption{Combined transition redshift (zt) from H(z), NVSS-ISW and Union 2.1 for flat Lambda-CDM model}
		\begin{tabular}{ll}
			\toprule
			\textbf{Datasets} & \textbf{Tranision Redshift (zt)} \\
			\midrule
			H(z)+Union 2.1 & 0.712±0.051 \\
			H(z)+Union 2.1+NVSS-ISW & 0.724±0.047 \\
		\end{tabular}%
		\label{tab:table_5}%
	\end{table}%

	\section{Conclusion}\label{conclusion}
	
	In this study, we started with an introduction to the deceleration parameter (q\textsubscript{0}) and then estimated it directly using the cosmic chronometer or H(z) data, and in the later parts of the chapter, we estimated q\textsubscript{0} by deriving it from the cosmological parameter results obtained using the cosmic chronometers data, type Ia supernovae data and then ISW data. We obtained the combined q\textsubscript{0}=-0.5808±0.025 for Lambda-CDM model,   and q\textsubscript{0}=-0.61±0.15 for both WCDM and CPL models. We also measured the jerk parameter (j\textsubscript{0}). The best fit measurement we got for the combined jerk parameter for Lambda-CDM model is 1±3.971e-07, for WCDM model is 1.054±0.141 and for CPL model is 1.0654±0.1345. We also estimated the transition redshift. We measured the combined zt=0.724±0.047 from the Lambda-CDM model parameters.

\end{document}